\documentclass[aps,prr,twocolumn,showpacs,floatfix,superscriptaddress,longbibliography]{revtex4-1}
\pdfoutput=1 

\usepackage{amssymb,amsmath,amstext} 
\usepackage{graphicx}                
\usepackage{epstopdf}                
\usepackage{color}                   
\usepackage{bm}                      
\usepackage{appendix}                
\usepackage[utf8]{inputenc}
\usepackage{bbold}
\usepackage{bbm}   
\usepackage{ulem}
\usepackage{latexsym}
\usepackage[colorlinks=true,allcolors=blue]{hyperref}
\usepackage{mathtools}
\usepackage[T1]{fontenc}

\usepackage[euler]{textgreek}
\usepackage{upgreek}

\usepackage[dvipsnames]{xcolor}
\definecolor{dgreen} {RGB}{0,100,0}

\renewcommand{\vec}[1]{\mathbf{#1}}

\allowdisplaybreaks

\begin{document} 

\title{Drag-induced dynamical formation of dark solitons in Bose mixture on a ring}
\author{Andrzej Syrwid} 
 \affiliation{Department of Physics, KTH Royal Institute of Technology, Stockholm SE-10691, Sweden}

 \author{Emil Blomquist}
 \affiliation{Department of Physics, KTH Royal Institute of Technology, Stockholm SE-10691, Sweden}

 \author{Egor Babaev}
 \affiliation{Department of Physics, KTH Royal Institute of Technology, Stockholm SE-10691, Sweden}

\begin{abstract}
Andreev-Bashkin drag plays a very important role in multiple areas like superfluid mixtures, superconductors and dense nuclear matter. Here, we point out that the drag phenomenon can be also important in physics of solitons, ubiquitous objects arising in a wide array of fields ranging from tsunami waves and fiber-optic communication to biological systems. So far, fruitful studies were conducted in ultracold atomic systems where nontrivial soliton dynamics occurred due to inter-component density-density interaction. In this work we show that current-current coupling between components (Andreev-Bashkin drag) can lead to a substantially different kind of effects, unsupported by density-density interactions, such as a drag-induced dark soliton generation. This also points out that soliton dynamics can be used as a tool to experimentally study the dissipationless drag effect.
\end{abstract}


\maketitle

Solitons are ubiquitous objects appearing in various physical systems, including nonlinear optics, fluid dynamics~\cite{Hasegawa1989, Dauxois2006PhysicsOS, Kivshar2003, GrimshawSolitonsFluids, Hasegawa:02, Henderson99, Kibler2010, Bailung2011,Chabchoub2020} and ultracold atomic systems~\cite{SuperfluidPitaevskii2016,Pethick2002}.
Ultracold bosons form Bose--Einstein condensate (BEC) 
effectively described by the Gross--Pitaevskii equation (GPE)~\cite{SuperfluidPitaevskii2016,Pethick2002,svistunov2015superfluid}.
The nonlinearity present in GPE can balance dispersive effects, supporting nonuniform solutions (solitons) preserving shape in time.
This, together with a great progress in cold atoms experimental techniques, makes ultracold bosonic systems an excellent platform for the studies on matter-wave solitons~\cite{Denschlag2000, Strecker2002, Khaykovich2002, Becker2008, Stellmer2008, Weller2008, Theocharis2010, Gawryluk_2005, Burger99, Carr2001,efimkin2016non,nguyen2014collisions,aycock2017brownian,fritsch2020creating,hurst2017kinetic}.  
Solitons also occur in fermionic ultracold atomic systems
\cite{Karpiuk_2002FermiSol,dziarmagasoliton2004,AntezzaSolitonFermi2007,SachaDelande2014ProperPhase,efimkin2015moving}. 

A conventional  superfluid is described by a complex field
$\psi=\sqrt{n}\, \mathrm{e}^{\mathrm{i} \varphi}$. 
The phase gradient can be identified with the superfluid velocity $\vec v=\frac{\hbar}{m}\nabla \varphi$, where $m$ is the particle mass~\cite{Onsager1949,
SuperfluidPitaevskii2016,Pethick2002,svistunov2015superfluid}.
In 1976 Andreev and Bashkin demonstrated that in a two-component interacting superfluid mixture
the relation between superfluid velocities and superflows becomes very nontrivial due to existence of a dissipationless drag transport effect~\cite{andreev1976three}. 
Indeed, the corresponding free-energy density takes the form $f= \sum_\alpha \rho_\alpha \vec v_\alpha^2/2 +\rho_\mathrm{d}\vec v_a \cdot \vec{v}_b$, where $\rho_\alpha$ ($\vec v_\alpha$) represents a superfluid density (superfluid velocity) of component $\alpha\in\{a,b\}$, and $\rho_\mathrm{d}$  is the Andreev--Bashkin (AB) drag coefficient~\cite{andreev1976three}. 
Consequently, the superflows, i.e., $\vec j_{\alpha}=\partial_{\vec v_\alpha}f
=\rho_\alpha \vec v_\alpha +\rho_\mathrm{d} \vec v_{\beta\neq\alpha} $, reveal that the component possessing no superfluid velocity, e.g., $\vec v_a= \vec 0$,  will still exhibit a nonzero superflow, $\vec j_a\neq\vec 0$,  as long as $\vec v_b \neq\vec 0$.

The AB effect strongly affects vortex lattices in superfluids~\cite{Dahl2008hidden,dahl2008unusual} and can change the nature of topological solitons in superconducotrs~\cite{RybakovHopf-Skyrme2019}. 
It is also crucial for the understanding of properties of dense nuclear matter~\cite{Sjoberg1976LandauEffMassNuclearMatter1,Chamel} and observed pulsar dynamics~\cite{alpar1984rapid,Alford2008FluxTubes,babaev2009unconventional}.
At the microscopic level the drag effect originates from inter-component particle-particle interaction \cite{andreev1976three,Fil2005Drag,linder2009calculation, hofer2012superfluid, hartman2018superfluid, colussi2021quantum}.
Especially interesting is the case of strongly correlated superfluids which parameters are precisely controllable in optical lattices~\cite{greiner2002quantum, bloch2005ultracold}. There the AB drag originates from the interplay between inter-component particle-particle interaction and lattice effects and can be, in relative terms, arbitrarily strong and $\rho_\mathrm{d}$ can be also negative~\cite{Kuklov2003counterflow,kuklov2004commensurate,kuklov2004superfluid,Fil2005Drag,capogrosso2008monte,dahl2008preemptive,linder2009calculation,capogrosso2011superfluidity,hofer2012superfluid,sellin2018superfluid,blomquist2021borromean,contessi2021collisionless,Nespolo2017avdreev,hartman2018superfluid,colussi2021quantum,gremaud2021pairing,contessi2021collisionless,linder2009calculation}.
Interestingly, AB drag signatures have been found in quantum droplets collisions~\cite{PylakDropletAB}.
The drag effect can have various forms. Recently, it was  demonstrated that in certain asymmetrical lattices there exists also a perpendicular entrainment referred to as vector drag~\cite{vecdrag}.

In binary systems very interesting solitonic effects are driven by inter-component density-density interaction ~\cite{Ohberg2001, Busch2001,Karpiuk2004bosefermi,Kevrekidis2004, Brazhnyi2005,GubeskysGapSolitons2006,Karpiuk2006bosefermi,Susanto2007,Doktorov2008,Rajendran_2009,Niederberger2010,Hoefer2011,Li_2015,Katsimiga2017,Farolfi2020, Cheng2006,Grimshaw2020,Arazo2021,Cidrim_2021}.  
In this paper, we study the consequences  of the AB effect (current-current interaction) on the solitonic dynamics. 
We consider a one-dimensional (1D) binary bosonic superfluid mixture modelled by the energy functional
$\mathcal{E}=N\int (\varepsilon_0+\varepsilon_\mathrm{d}) \,  \mathrm{d}x$ 
with 
$\varepsilon_0=\sum_\alpha[-\hbar^2\psi^*_\alpha \partial_x^2\psi_\alpha/2m +\mathrm{g}_{\alpha} N|\psi_\alpha|^4/2]$
and $\varepsilon_\mathrm{d}=\mathrm{g}_\mathrm{d}N\sum_\alpha J_\alpha^2/2+\mathrm{g}_\mathrm{d} N J_a J_b=\mathrm{g}_\mathrm{d}N(J_a+ J_b)^2/2.$
Here $\psi_\alpha$ is the condensate field of component $\alpha\in\{a,b\}$ normalized to unity $|\langle \psi_\alpha |\psi_\alpha\rangle|^2=1$.
The particles, which numbers are equal and conserved in both components, $N_\alpha=N$, possess equal massees $m_\alpha=m$ and are confined in a ring of circumference $L$, i.e., we assume periodic boundary conditions (PBC) $\psi_\alpha(x+L,t)=\psi_\alpha(x,t)$.
The condensates are subjected to an intra-component contact interaction of strength governed by $ g_{\alpha0} $
and the AB inter-component drag incorporated by scalar product of  
$J_\alpha=\hbar\psi_\alpha^* \partial_x \psi_\alpha/(2m \mathrm{i})+\text{c.c.}$
with strength given by $\mathrm{g}_\mathrm{d}>0$.
The contributions $\propto  J _\alpha^2$ in $\varepsilon_\mathrm{d}$ are required for  $\mathcal{E}$ to be bounded from below.
Such a phenomenological effective model of the AB drag has previously been studied in other contexts~\cite{GaraudSkyrmions2014,RybakovHopf-Skyrme2019}.

\begin{figure}[t!]
  \includegraphics[width=1\columnwidth]{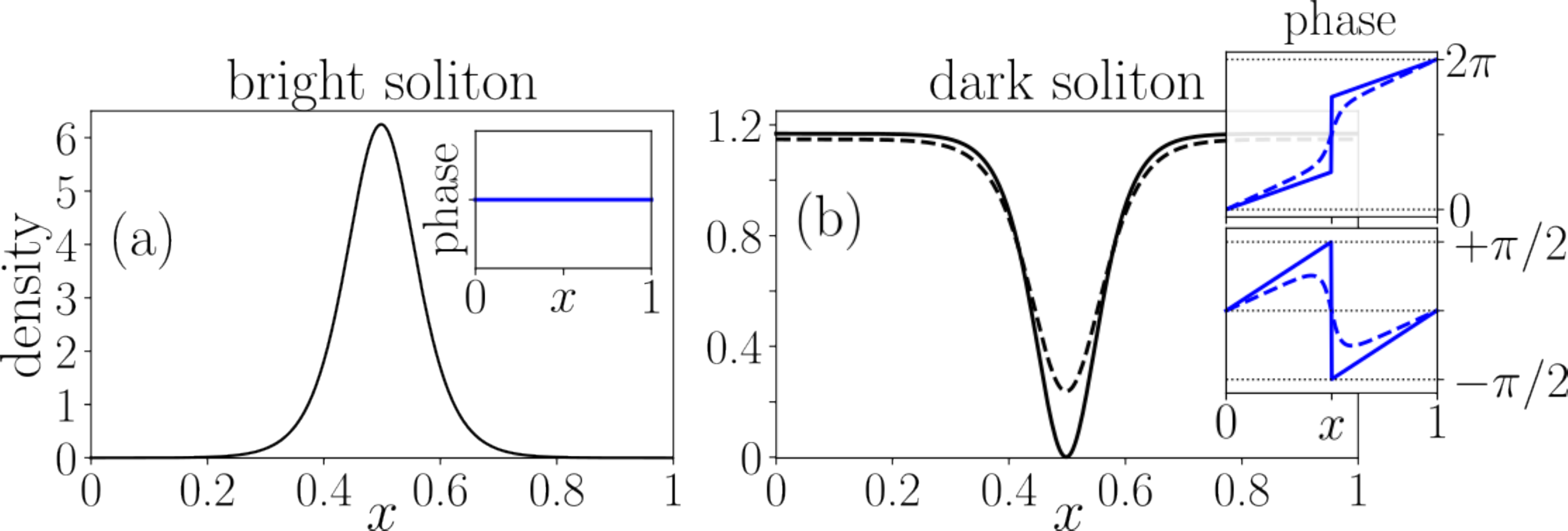}
  \caption{Illustration of well localized solitons confined in a ring of circumference $L=1$. Panel (a) shows a lowest energy bright soliton density while panel (b) presents densities of two types of dark solitons: black (solid line) and gray (dashed line). The corresponding phase distributions are depicted in the respective insets. While the stationary bright soliton in (a) has a uniform phase, a dark soliton notch is always accompanied by a phase slip that can be of two kinds: facing down or facing up. Upper and lower insets of (b) show phase distributions characterized by $W=1$ and $W=0$, respectively. 
  }\label{f1}
\end{figure}

Our goal is to investigate the effects of current-current interaction. Hence, in this work we specifically set the well-studied
inter-component density-density interaction to zero. 
However, the effect of the latter is  discussed in Supplemental Material (SM)~\cite{SM}.
The corresponding system of dimensionless time-dependent GP-like equations reads ($\alpha\in\{a,b\}$, $\gamma \neq\alpha$) 
\begin{align}
    \mathrm{i}\partial_{t}\psi_\alpha = 
    -\frac{\partial_x^2\psi_\alpha}{2}
    +g_\alpha|\psi_\alpha|^2\psi_\alpha
    +g_\mathrm{d}\mathcal{J}_{\alpha\alpha}
    +g_\mathrm{d}\mathcal{J}_{\alpha\gamma},
    \label{gpe1}
\end{align} 
where the length scales are measured in units of the ring circumference $L$, 
time in units of $\frac{mL^2}{\hbar}$, 
energy in units of $\frac{\hbar^2}{mL^2}$,
and we defined $g_\alpha = \frac{mL}{\hbar^2}\mathrm{g}_{\alpha}N$, $g_\text{d} = \frac{N}{mL}\mathrm{g}_{\text{d}}$, and $\mathcal{J}_{\alpha\beta}=[2 (\partial_x \psi_\alpha)J_\beta+\psi_\alpha (\partial_x J_\beta)]/2\mathrm{i}$ with the dimensionless $J_\alpha = \psi_\alpha^* \partial_x \psi_\alpha/2\mathrm{i}+\text{c.c.}$.
In the absence of drag, i.e., for $g_\mathrm{d}=0$, Eqs.~(\ref{gpe1})
become independent and support both bright and dark soliton solutions that for PBC can be expressed analytically in terms of Jacobi functions~\cite{CarrRepulsive2000,CarrAttractive2000,KanamotoMetastable2009,WuZaremba2013,SyrwidTutorial2021}.
A stationary bright soliton in ring geometry (PBC)
forms spontaneously in the ground state when $g_\alpha<g_c=-
\pi^2$. On the other hand, dark solitons are collective excitations characterized by density notches accompanied by phase slips in phase distribution $\varphi$ and appear for any $g_\alpha>0$~\cite{KanamotoPhaseTransition2003,SyrwidTutorial2021}. 
For finite rings, i.e., $L<\infty$, a single dark soliton always propagates with some finite velocity because the phase cyclicity condition, $\varphi(L)-\varphi(0)=2\pi W$ where the winding number $W\in \mathbb{Z}$, requires a nonzero phase gradient to be satisfied in the presence of a solitonic phase slip.
In the limiting case of a totally dark, i.e., {\it black},  soliton the corresponding density vanishes in the dip where the phase reveals a single-point discontinuity by $\pi$. Therefore, to satisfy PBC the phase $\varphi$ has to accumulate at least as $ \pm i \pi x/L $. Solitons with a shallower density notch accompanied by a smooth $\varphi(x)$ are often called {\it gray} solitons. Note that two gray solitons revealing identical densities may possess phase distributions characterized by different $W$ and in consequence different average momenta $\langle p\rangle =-\mathrm{i}\hbar\int\mathrm{d}x \, \psi^* \partial_x \psi$. In Fig.~\ref{f1} we show typical density and phase distributions of the lowest energy bright soliton and two types of dark solitons: black and gray. 

From the many-body perspective, dark solitons are 
directly connected with a specific class of the so-called yrast states~\cite{Kulish76,Ishikawa80,Kanamoto2008,Kanamoto2010,Komineas2002,Jackson2002,Sato2012,Karpiuk2012,Karpiuk2015,Syrwid2015,Sato_2016,Syrwid2016,Gawryluk2017,Oldziejewski2018,Shamailov2019,Golletz2020,Shamailov_2016,Syrwid2018YangGaudin}, i.e., lowest energy states for a given total momentum. Similar many-body excitations correspond to dark solitons also
in the presence of open boundary conditions~\cite{Syrwid2017HardWalls}. For an overview see~\cite{SyrwidTutorial2021}. Here, we study whether current-current drag interactions can lead to yrast excitations, inducing formation of dark solitons.

Let us assume that in our system
one of the components,  say the $b$-component, exhibits 
$J_b\neq0$ while $J_a=0$. If the spatial translation symmetry is broken and $\mathcal{J}_{\alpha\gamma}\neq0$,
then a dynamic drag-related current generation and a momentum transfer between the components can be expected.
To study this problem, we consider the case in which component $a$ is initially prepared 
in the uniform ground state $\psi_{a0}$ for repulsive interaction $g_a>0$.
At the same time $b$ component is prepared in the ground state, $\psi_{b0}$, but for attractive interactions characterized by $g_b<g_c$ that is associated with a stationary bright soliton.
In such a case $\langle p_a \rangle =\langle p_b \rangle =0$, $J_a=J_b=0$ and the drag interactions have no impact on these states.
To have $J_b,\mathcal{J}_{\alpha\gamma}\neq0$ we additionally set the bright soliton in motion such that initially $\langle p_b \rangle, J_b \neq0$.

Basing on the relationship between yrast states and dark solitons in a single component repulsive Bose gas with PBC, one can ask if the drag-related momentum transfer from component $b$ to $a$ can induce a dark soliton formation in the latter component. 
We argue that preparing component $ b $
in a well localized bright soliton state may  reduce excitations of kinds other than the collective solitonic ones. That is, the bright soliton would slow down its propagation when transferring the momentum from $b$ to $a$, while preserving approximately unchanged shape due to strong intra-component attraction.
In such a case, there is a chance that most of the energy gained by component $a$ would correspond to the collective motion characterized by the transferred momentum. Thus, excluding the drag interaction energy, the resulting excited state in $a$ component would have energy close to the one possessed by the yrast state with $\langle p_a\rangle$.
If so, then one may expect an emergence of dark soliton signatures (density notch and phase slip) in the $a$-component.

Given that the abovementioned scenario takes place, the induced dark soliton is expected to be different depending on the amount of momentum injected into component $ a $---the latter is likely to change over time.
One may ask whether or not it is possible
for a specific dark soliton to form in component $ a $ that would coexist with the bright soliton in the other component for time-scales longer than the period of a single revolution of the anticipated dark soliton along the ring.
We suppose that this can happen when both the target dark soliton and the bright soliton propagate with comparable velocities.

The well localized (narrow in comparison to $L$) bright soliton can be approximately described by the famous sech-shaped soliton wave function~\cite{Pethick2002} which reveals its particle-like behavior. Note that $\langle p\rangle =  \hbar \int \mathrm{d}x |\psi|^2 \partial_x \varphi$ and $\varphi(x)=\varphi(0)+m v x/\hbar +\mathcal{S}(x)$, where $\mathcal{S}(x)$ encodes other phase features like phase slips. For well localized bright solitons $\partial_x \mathcal{S}\approx 0$ in the vicinity of the soliton clump and thus such states propagate with the velocity $v\approx\langle p \rangle/m$. 
Generally, $\int \mathrm{d}x|\psi|^2\partial_x \mathcal{S}$ is  nonnegligible for dark solitons making the relationship between $v$ and $\langle p \rangle$ more complicated. The special case is a black soliton (bs), for which $\partial_x \mathcal{S}\neq0$ only at the soliton dip where $|\psi_\text{bs}|^2=0$.
Thus, for the black  soliton $v_\text{bs}=\langle p_\text{bs} \rangle/m$, where $\langle p_\text{bs} \rangle/\hbar= \pi/L+2\pi n/L$ with $n\in \mathbb{Z}$.

Let us operate with the dimensionless units and restrict our considerations to
states $\psi_\alpha$ possessing $0\leq\langle p_\alpha \rangle\leq 2\pi$ measured in $\hbar/L$ units. We are going to analyze the possibility of a drag-induced formation of the most distinct of dark solitons, namely, the black soliton. We suppose that a long living coexistence of black and bright solitons may be possible when both objects propagate with comparable velocities. Therefore, at $t=0$, we set the initial ground state bright soliton ($b$ component) in motion with $\langle p_b\rangle=2\langle p_\text{bs}\rangle =2\pi$. This is done by multiplying $\psi_{b0}(x)$ by $\mathrm{e}^{\mathrm{i} 2\pi x}$, i.e., $\psi_b(x,t=0)=\psi_{b0}(x)\mathrm{e}^{\mathrm{i} 2\pi x}$. Since $\langle p_b\rangle +\langle p_a \rangle=2\pi$ is a conserved quantity in our system, we expect that if the momentum is transferred from component $b$ to $a$, the abovementioned coexistence my appear when $\langle p_b \rangle-\langle p_a \rangle \approx 0$. In such a case  $\langle p_b \rangle \approx \langle p_a \rangle \approx \pi$ and the corresponding solitons should propagate with comparable velocities. 

We prepared the initial bright soliton state $\psi_{b0}(x)$ by means of an imaginary time evolution of (\ref{gpe1}) with $\alpha=b$, $g_\mathrm{d}=0$ and four different $g_b=-20,-25,-30,-35$, separately. These values of $g_b$ are all substantially below the critical value $g_c=-\pi^2$ which guarantees that the resulting bright soliton density is well localized. This state is then set in motion with $\langle p_b\rangle|_{t=0}=2\pi$ by incorporating a phase factor as previously described. Component $a$ is prepared in a similar way but with $g_a\in\{20,25,\ldots,90\}$  resulting in the lowest energy state $\psi_{a0}= \psi_a(x,t=0)=1$ (up to a global phase). After the states preparation we switch on the AB drag by setting $g_\mathrm{d}=0.1$ while keeping $g_a$ and $g_b$ fixed. We then numerically evolve Eqs.~(\ref{gpe1}) in real time up to $t=10$, a time more than 30 times longer than the characteristic period of the black soliton revolution around the ring $T=1/\pi\approx 0.32$.

\begin{figure}[t!]
  \includegraphics[width=1\columnwidth]{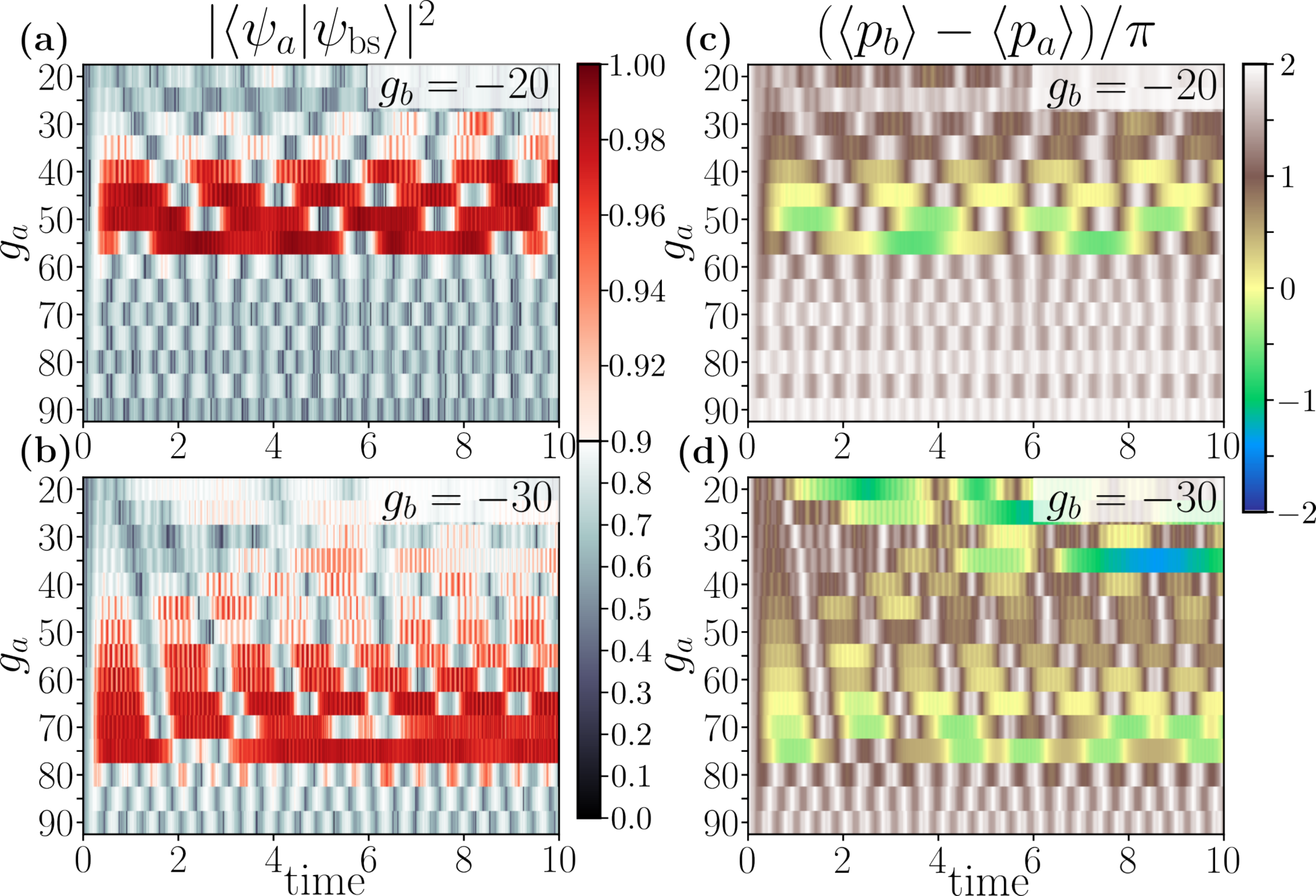}
  \caption{ Time evolution of the overlap $|\langle\psi_a|\psi_\mathrm{bs}\rangle|^2$ (left column) and  momentum difference $(\langle p_b\rangle-\langle p_a\rangle)/\pi$ (right column) obtained for $g_\mathrm{d}=0.1$.
  Consecutive rows
  correspond to different $g_b=-20,-30$ characterizing the bright soliton state. The regions where $|\langle\psi_a|\psi_\mathrm{bs}\rangle|^2>0.9$ coincide with small values of $|\langle p_b\rangle-\langle p_a\rangle |$ indicating that (nearly) black soliton forms when approximately half of  $\langle p_b \rangle|_{t=0}=2\pi$ is transferred to component $a$. 
  }\label{f2}
\end{figure}
Our results indicate that the bright soliton in $b$ component survives the evolution for all the considered parameters. For each $g_b=-20,-25,-30,-35$ we find 
a region in the $g_a$ parameter where clear dark soliton signatures (density notch and phase slip) emerge in $\psi_a(x,t)$.
See SM~\cite{SM} for snapshots of typical system dynamics. Fig.~\ref{f2} shows the temporal behavior of the overlap $|\langle \psi_a|\psi_\text{bs} \rangle|^2$  and 
the momentum difference $(\langle p_b\rangle -\langle p_a\rangle)/\pi$ for different $g_a$ and $g_b=-20, -30$. The overlaps $|\langle \psi_a|\psi_\text{bs} \rangle|^2$ are calculated with the analytical black soliton solution $\psi_\text{bs}$ characterized by the corresponding $g_a$ and located at a position of the phase slip recognized in $\psi_a(x,t)$.
By choosing a specific color code in the overlap plots we discriminate the regions where $|\langle \psi_a|\psi_\text{bs} \rangle|^2>0.9$ (red intensity) from those where $|\langle \psi_a|\psi_\text{bs} \rangle|^2<0.9$ (gray intensity). Note that overlaps above 0.9 appear when the momenta $\langle p_b\rangle$ and $\langle p_a\rangle$ are similar and are maintained for time scales significantly longer than $T$. We observe that the critical $g_a$ above which a dark soliton appears depends on the value of $g_b$. That is, for stronger attraction, i.e., narrower bright soliton in the $b$ component,
the regime of the (nearly) black soliton formation shifts to larger $g_a$'s corresponding to the narrower dark solitons. In SM [83] we also analyze how drag-induced states $\psi_a$  would evolve if drag is quenched to zero (drag-free dynamics) at a time when $|\langle \psi_a| \psi_\text{bs}\rangle|^2\approx 1$. It turns out that such generated states reveal a genuine dark soliton drag-free evolution.

\begin{figure}[t!]
  \includegraphics[width=1\columnwidth]{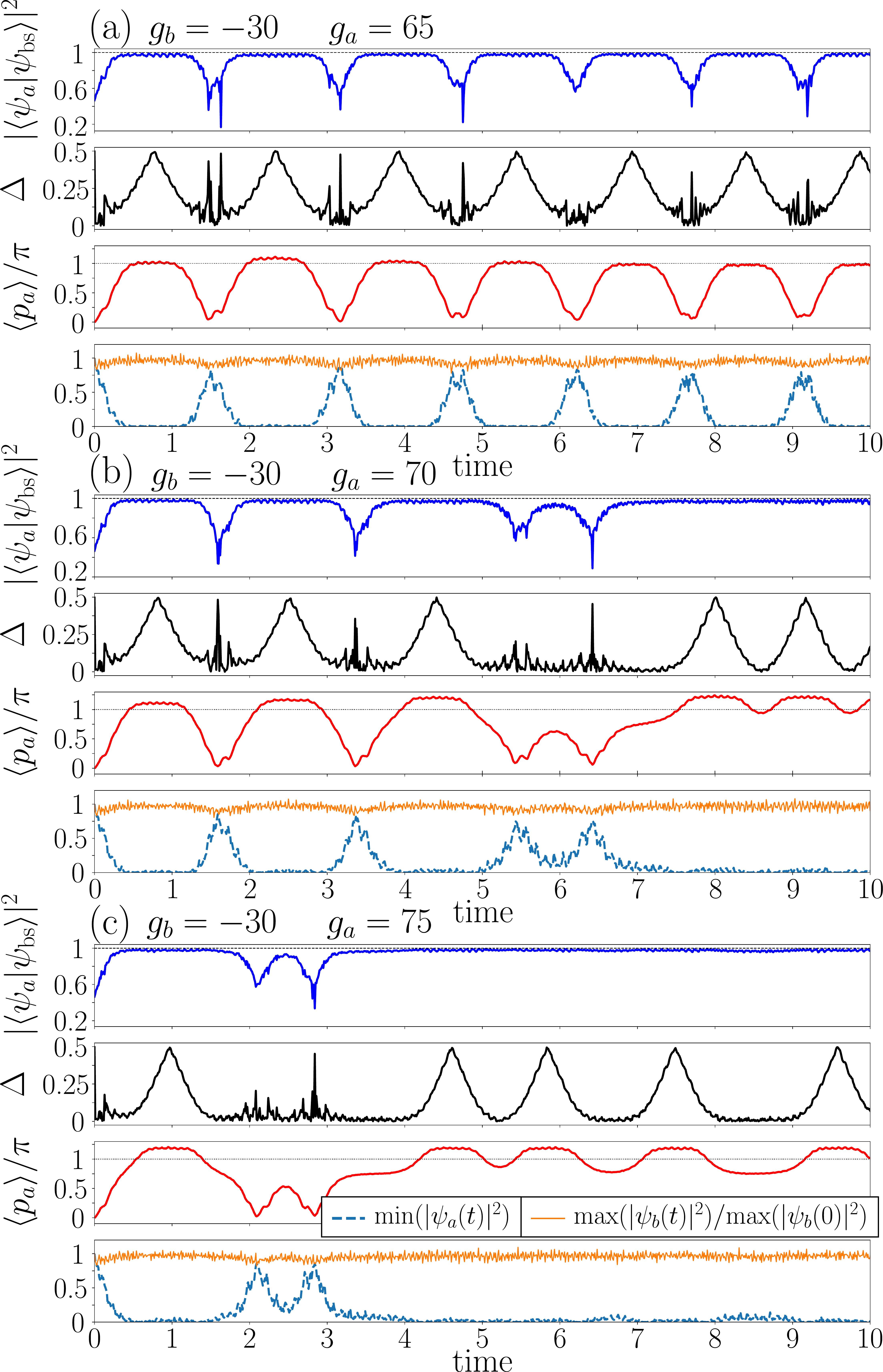}
  \caption{Each set of plots (a), (b), and (c) shows from top to bottom the dynamics of: the overlap $|\langle \psi_a|\psi_\mathrm{bs} \rangle|^2$, the relative distance $\Delta$ along the ring between the bright soliton ($b$ component) and the phase slip position in $\psi_a(x,t)$, the average momentum $\langle p_a\rangle/\pi$, as well as the values $\mathrm{min}(|\psi_a(t)|^2)$ and $\mathrm{max}(|\psi_b(t)|^2)/\mathrm{max}(|\psi_b(0)|^2)$. The results in (a), (b), and (c) correspond to $g_b=-30$ and $g_a=65,70,75$, respectively. The drag-induced dark (nearly black) soliton  often is significantly disturbed, or even completely destroyed, when passing through the bright soliton, i.e., when $\Delta \rightarrow 0$. In such a case the phase slip in $\psi_a(x,t)$ is rather tiny or even unrelated to any soliton structure.  This is the origin of the narrow spikes observed in the $\Delta$ plots when $\Delta\rightarrow 0$ and  $\mathrm{min}(|\psi_a(t)|^2) \approx 1$. Nevertheless, as shown in (b) for $t>7$ and in (c) for $t>3$, the (nearly) black soliton can survive encounter with the bright soliton. 
  }\label{f3}
\end{figure}

To better understand the system dynamics, in Fig.~\ref{f3} we closer study cases with $g_b=-30$ and $g_a=65,70,75$. 
As before, we analyze the time dependence of the overlap $|\langle\psi_a|\psi_\mathrm{bs}\rangle|^2$ and momentum $\langle p_a \rangle/\pi$. Additionally, we monitor the minimum Euclidean distance $\Delta$ along the ring  between the bright soliton and the drag-induced dark soliton, the minimum reached by an anticipated density notch $\mathrm{min}(|\psi_a(t)|^2)$, as well as the ratio of the bright soliton height to its initial value $\mathrm{max}(|\psi_b(t)|^2)/\mathrm{max}(|\psi_b(0)|^2)$. In all the cases an initial momentum transfer leads to the formation of a (nearly) black soliton. Indeed, the overlap $|\langle\psi_a|\psi_\mathrm{bs}\rangle|^2$ increases together with $\langle p_a \rangle$, and the density notch is simultaneously being carved as indicated by the decreasing value of $\mathrm{min}(|\psi_a(t)|^2)$. At the same time the distance $\Delta$ reveals an increasing separation between solitons in the two components reaching maximum, $\Delta\approx 0.5$, at a time in the middle of the plateau of
 $ |\langle\psi_a|\psi_\mathrm{bs}\rangle|^2 \approx 1 $.
The seemingly linear trend in $ \Delta $ for 
$ \Delta \gtrsim 0.1 $ reveals a constant relative motion between the spatially separated solitons $|v_b-v_a| \approx 1$ three times slower than the single-component black soliton velocity $v_\text{bs}=\pi$. This behavior of $\Delta$ repeats multiple times during the evolution.

Due to different velocities and assumed ring system geometry, the solitons collide multiple times during the course of evolution. It turns out that the induced (nearly) black soliton state often is substantially disturbed or even completely destroyed when both solitons meet, i.e., when $\Delta\rightarrow0$, which results in an abrupt drop of the overlap value $|\langle\psi_a|\psi_\mathrm{bs}\rangle|^2$. The dark soliton re-localizes again when $\Delta$ increases. Such a mechanism is the origin of quasi-periodic patterns visible in Figs.~\ref{f2} and~\ref{f3}.
However, as indicated by the behavior of $\mathrm{max}(|\psi_b(t)|^2)/\mathrm{max}(|\psi_b(0)|^2)$, the bright soliton remains almost unaffected when passing through the dark one. On the other hand, as shown in Fig.~\ref{f3}(b) for $t>7$ and Fig.~\ref{f3}(c) for $t>3$, the drag-induced dark soliton can also survive an encounter with the bright soliton. Additionally, in  Fig.~\ref{f3}(c) for $t\in(6.3,7)$ and $t\in(8,9)$, one can observe signatures of the existence of long living dark-bright soliton composites characterized by $\Delta \approx 0$. 
For more intuition, see snapshots of the system evolution in SM~\cite{SM}.

In summary, we have studied the dynamics of a bosonic binary mixture confined in a 1D ring geometry with intra-component contact interactions and inter-component Andreev--Bashkin drag. Based on the relationship between dark solitons and yrast states characterized by the lowest energy for a given momentum, we formulated and verified the hypothesis concerning a drag-induced dark soliton formation process. By numerically computing the system dynamics we tested the scenario where a propagating bright soliton interacts with the other component, prepared in the repulsively interacting
uniform ground state. 
We demonstrated that there exist parameter regimes for which the drag interaction leads 
to formation of a long living genuine, nearly black, soliton state in the initially uniform component.
While we focused on the most distinct black soliton case, the general idea provided here should also allow for generation of gray solitons. 
Our goal here was to study the effects of current-current interaction on soliton dynamics. 
An interesting question that warrants further studies is how these effects combine with inter-component density-density interactions. This question is beyond the scope of this paper, but in~\cite{SM} we show that the drag phenomenon is crucial for the dynamical formation of long-living dark solitons, while density-density inter-component coupling does not support this effect in the considered setup. Additionally, we present that the effect at least survives inclusion of not too strong density-density interactions. The discussed phenomenon could guide experiments for a detection of the AB drag effect in binary superfluids. This can open avenue of studying the drag effect directly in a laboratory shedding light on the drag effect in other systems ranging from multicomponent superconductors to superfluids in neutron stars.

In conclusion, previously, soliton physics in binary systems were restricted to the role of density-density  interaction.
In this paper we report that new kind of soliton dynamics arises in binary system due to current-current coupling. The results indicate that the mixed gradient coupling plays an important role in soliton physics in multi-component systems which warrants further investigation.  
We expect that competition between the   drag effect and density-density inter-component interactions leads to even richer dynamics of multicomponent systems.

\section*{Acknowledgements}
The authors are grateful to Krzysztof Sacha for valuable discussions.
E. Bl. and E. Ba. were supported by the Swedish Research Council Grants No. 2016-06122, 2018-03659, and G\"{o}ran Gustafsson Foundation for Research in Natural Sciences.
A. S. and E. Ba. acknowledge the support from Olle Engkvists stiftelse.

\bibliography{main}

\renewcommand{\thefigure}{FS\arabic{figure}}

\renewcommand{\theequation}{S\arabic{equation}}

\setcounter{figure}{0}
\setcounter{equation}{0}

\begin{titlepage}
\begin{center}
    \Large {\bf Supplemental Material}
\end{center}
    \vspace{1.cm}
\end{titlepage}

\section{Monitoring the system dynamics}
\vspace{-0.3cm}

\begin{figure*}[h!]
    \includegraphics[width=1.\textwidth]{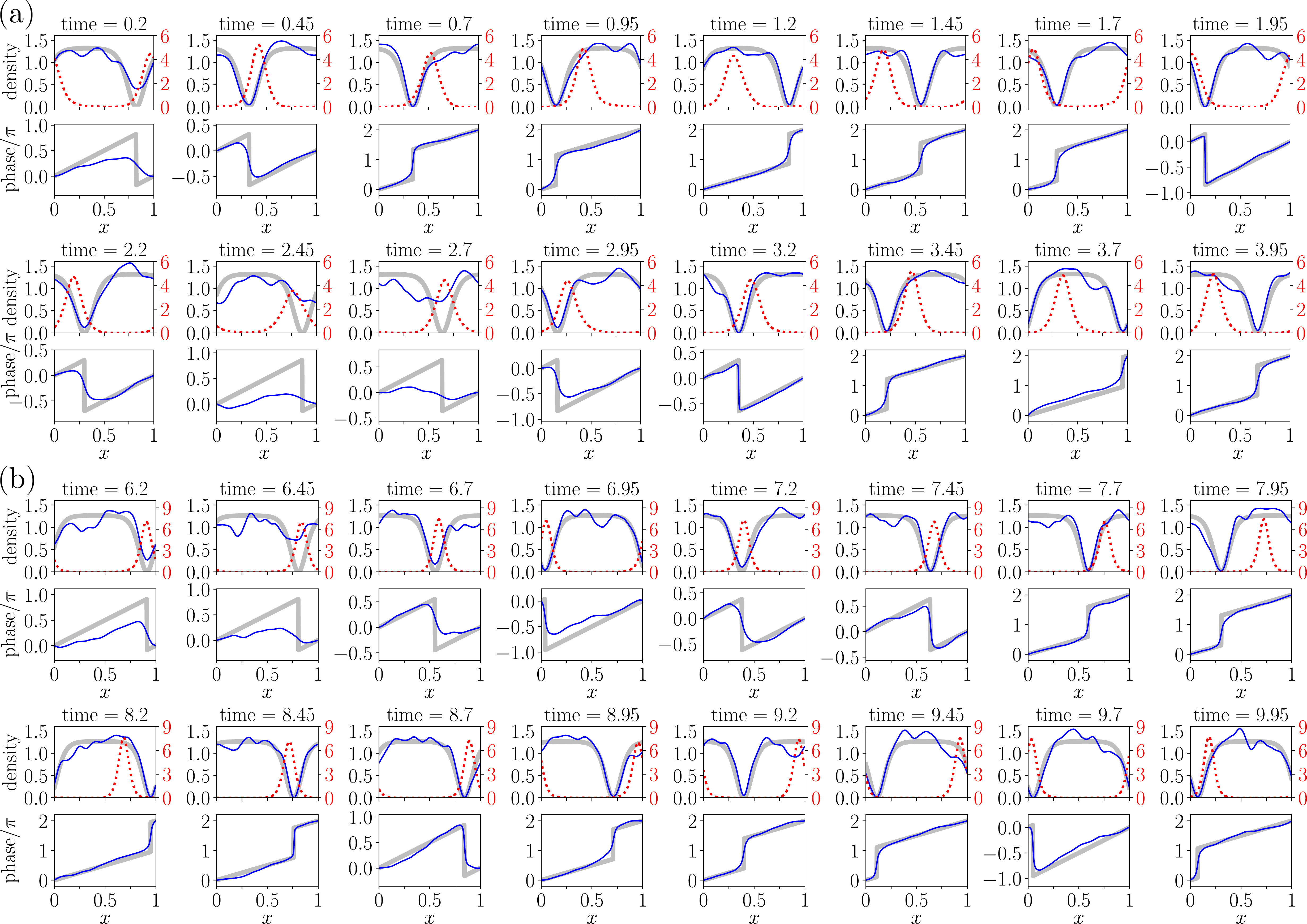}
\vspace{-0.5cm}
    \caption{(a): Snapshots of the system dynamics in the time frame $[0.2,3.95]$, obtained for $g_a=50$, $g_b=-20$, and $g_\text{d}=0.1$. The density and phase distribution of component $a$ (blue lines) are compared with the density and phase of the corresponding black soliton state $\psi_\text{bs}$ (gray lines). We also show the bright soliton density in $b$ (dotted red curves), where the corresponding scale is depicted on the right-hand side of the plots with red ticks.
    (b): Similar results as in (a) but obtained for $g_a=70$, $g_b=-30$, $g_\text{d}=0.1$ and in the time frame $[6.2,9.95]$. 
    }
    \label{snapshots}
\end{figure*}

In Fig.~\ref{snapshots} we present snapshots of  representative system dynamics in the setup described in the main text. We show two examples where the dark soliton state is induced in component $a$ due to the drag interactions with component $b$. That is, on the one hand in (a) the evolution of the system characterized by $g_a=50$, $g_b=-20$ and $g_\text{d}=0.1$ is monitored in the time frame from $t=0.2$ to $t=3.95$. On the 
other hand in (b) we track the dynamics for parameters $g_a=70$, $g_b=-30$, $g_\text{d}=0.1$ in time between $t=6.2$ and $9.95$. Note that the drag-induced dark soliton in component $a$ for most of the time is very similar to the corresponding black soliton state indicated in the plots for comparison. Nevertheless, it can be significantly disturbed or even completely disappear when passing through the bright soliton in the other component. 

\onecolumngrid
    
\twocolumngrid

\section{Drag-free dynamics of induced states}

\begin{figure}[t!]
\includegraphics[width=1.\columnwidth]{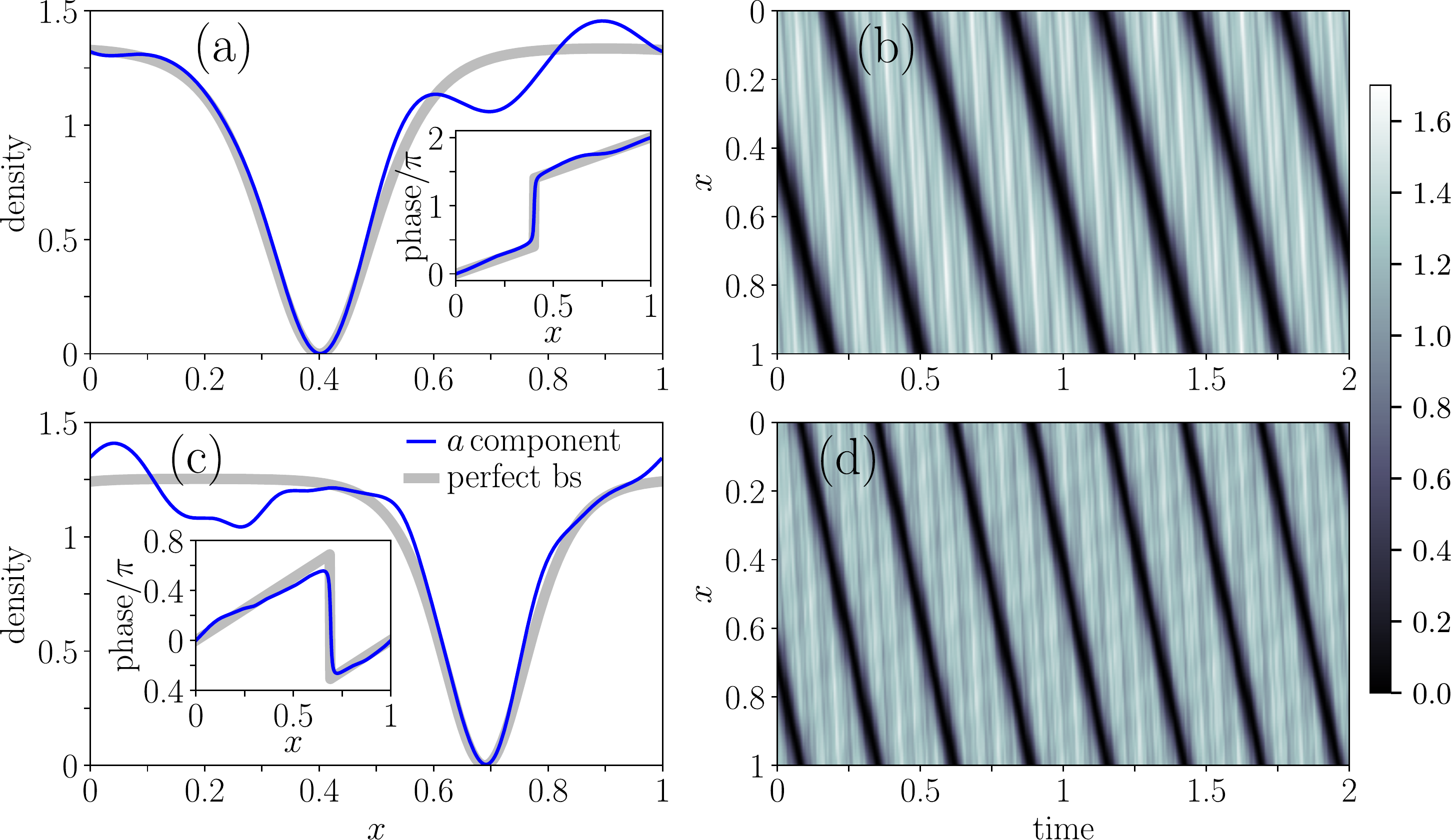}
\vspace{-.25cm}
\caption{
Panel (a) presents the density (blue line) and phase distribution (inset; blue line) of $\psi_a(x,t)$ after a time $t=1.0$ of evolution for $g_a=45$ and $g_\mathrm{d}=0.1$ with the bright soliton in component $b$ characterized by $g_b=-20$. For comparison we show  density and phase distribution of the corresponding black soliton state $\psi_\mathrm{bs}$ (gray lines). The state $\psi_a (x,t=1)$ is a representative example of states revealing a high overlap $|\langle\psi_a|\psi_\mathrm{bs}\rangle|^2$. Panel (b) shows the drag-free ($g_\mathrm{d}=0$) genuine dark soliton dynamics of the density (color code) of $\psi_a$ presented in (a). 
In (c) and (d) we present similar results but obtained for $t=7.0$, $g_a=75$, $g_b=-30$, and $g_\mathrm{d}=0.1$. 
}
\label{drag-free_evol}
\end{figure}

Let us look closer at typical drag-induced dark (nearly black) solitons generated in the $a$ component. The left panels of Fig.~\ref{drag-free_evol} present two representative examples of $\psi_a(x,t)$ obtained for $g_a=45$, $g_b=-20$ at $t=1.0$ (a) and for $g_a=75$, $g_b=-30$ at $t=7.0$ (c), respectively. The results reveal density notches and phase distributions very similar to the corresponding black soliton solutions. Here, we also verify the single-component drag-free dynamics of these drag-induced states in right panels of Fig.~\ref{drag-free_evol}. That is, by employing Eq.~(1) with $\alpha=a$ and $g_\mathrm{d}=0$ we performed the numerical evolution assuming that the initial states are represented by $\psi_a$'s illustrated in \ref{drag-free_evol}(a) and \ref{drag-free_evol}(c). The resulting time dependencies of corresponding densities shown in \ref{drag-free_evol}(b) and \ref{drag-free_evol}(d) reveal genuine dark soliton dynamics---in both cases the well preserved soliton notch propagates with a constant velocity. We would like to stress that also the phase flip signature of the dark soliton survives and always coincides with the corresponding density notch during the evolution.

\vspace{-0.25cm}

\section{Inclusion of density-density intercomponent interactions}

Here, we discuss the impact of density-density intercomponent interactions on the dynamical dark soliton formation effect. For this purpose we extend our model so that $\varepsilon_0+\varepsilon_\text{d}\rightarrow\varepsilon_0+\varepsilon_\text{d}+\varepsilon_{d-d}$, where $\varepsilon_{d-d}=\mathrm{g}_{ab}N|\psi_a|^2|\psi_b|^2$.
In consequence, the dimensionless GP-like equations (1) gain an additional term and read ($g_{ab}=\frac{m L}{\hbar^2}\mathrm{g}_{ab} N, \, \gamma\neq\alpha$)
\begin{align}
    \mathrm{i}\partial_{t}\psi_\alpha = 
    &-\frac{\partial_x^2\psi_\alpha}{2}
    +g_\alpha|\psi_\alpha|^2\psi_\alpha
    +g_{ab}|\psi_\gamma|^2\psi_\alpha
    \nonumber
    \\
    &+g_\mathrm{d}\mathcal{J}_{\alpha\alpha}
    +g_\mathrm{d}\mathcal{J}_{\alpha\gamma}
    .
    \label{inclgab_gpe}
\end{align}

\begin{figure}[h!]
\includegraphics[width=1.0\columnwidth]{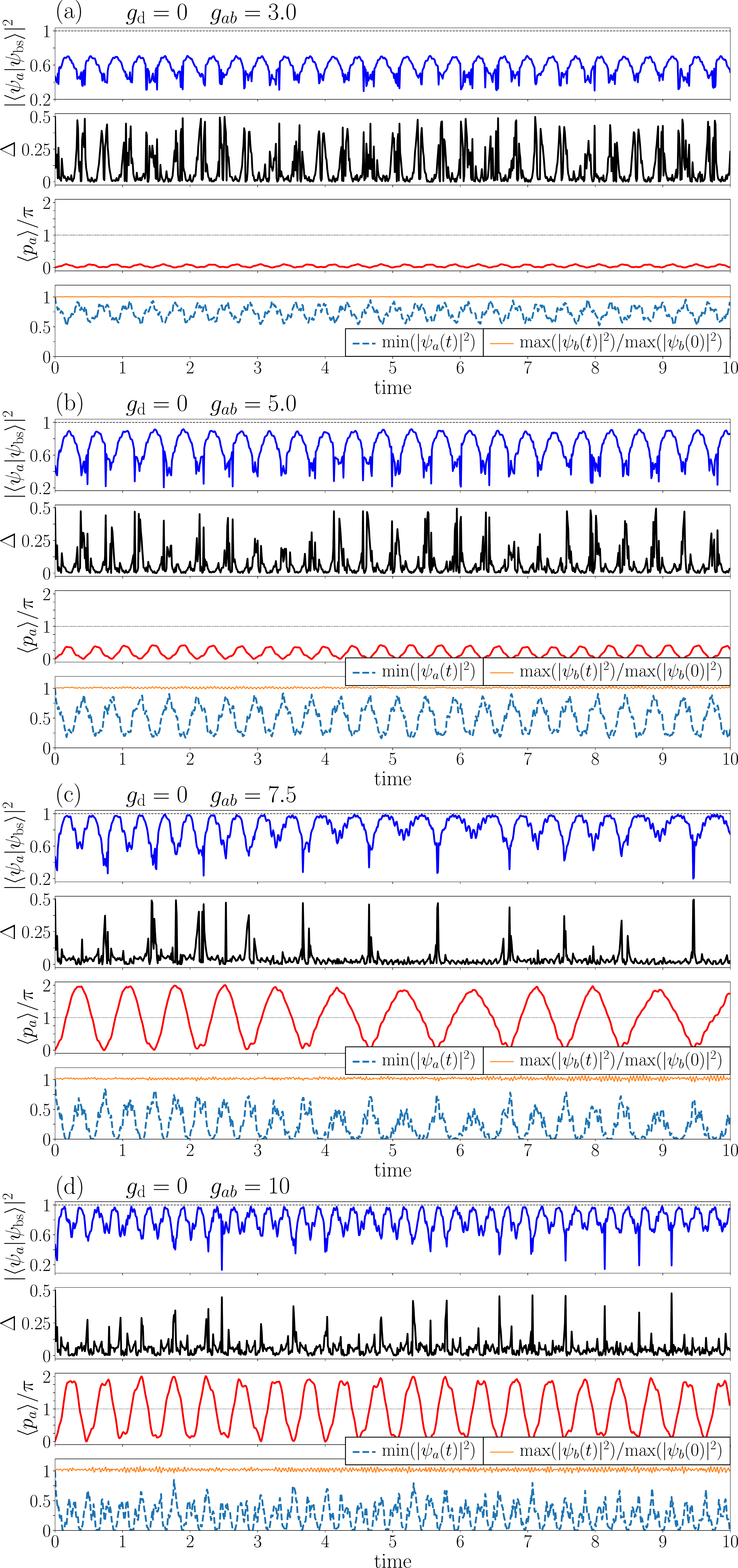}
\vspace{-0.5cm}
\caption{Dynamics with exclusively density-density intercomponent interactions ($g_\text{d}=0$, $g_{ab}\neq0$).
From top to bottom the time dependence of: the overlap $|\langle \psi_a|\psi_\mathrm{bs} \rangle|^2$, the relative distance $\Delta$ along the ring between the bright soliton in $b$ and a phase slip position in $a$ component, the average momentum $\langle p_a\rangle/\pi$, as well as the values $\mathrm{min}|\psi_a(t)|^2$ and $\mathrm{max}|\psi_b(t)|^2/\mathrm{max}|\psi_b(0)|^2$.
The results in (a), (b), (c), and (d) correspond to $g_b=-30$, $g_a=75$, $g_\text{d}=0$, and $g_{ab}=3,5,7.5,10$, respectively. 
Note that even if the overlap $|\langle \psi_a|\psi_\mathrm{bs} \rangle|^2$ reaches values close to 1, it is never maintained for a noticeable time. 
}
\label{excl-dd}
\end{figure}

Let us first consider the case of exclusively density-density intercomponent interactions $g_{ab}\neq0$ by setting $g_\mathrm{d}=0$. While the density-density interactions can also lead to a momentum transfer between components, they turn out to be insufficient for formation of a long living dark soliton in the setup analyzed in the main text. In Fig.~\ref{excl-dd} we present a few examples of the system dynamics for the parameters $g_a=75$, $g_b=-30$, $g_\mathrm{d}=0$ and different $g_{ab}\in\{3,5,7.5,10\}$. The results are organized as in Fig.~3 in the main text. Note that while the momentum transfer from $b$ to $a$ component coincides with an increase (and decrease) of the overlap $|\langle \psi_a | \psi_\mathrm{bs}\rangle|^2$, the system evolution does not reveal any noticeable time frames where $|\langle \psi_a | \psi_\mathrm{bs}\rangle|^2$, $\langle p_a\rangle/\pi$, and $\mathrm{min}(|\psi_a(t)|^2)$ are at least relatively stable, i.e., maintain approximately constant values. Therefore, despite the fact that purely density-density intercomponent interactions can lead to a dynamical formation of a dark soliton in the $a$ component, there is no stabilization mechanism and the soliton disappears very quickly, cf., very short periods of high overlaps  $|\langle \psi_a | \psi_\mathrm{bs}\rangle|^2$ and small values of $\mathrm{min}(|\psi_a(t)|^2)$ in panels (c) and (d) of Fig.~\ref{excl-dd}. Here, we would like to stress that the situation is even worse when dealing with intercomponent density-density attraction ($g_{ab}<0$). This is due to a tendency of $a$-component density accumulation around the bright soliton in $b$ component, instead of carving a density notch.   
In conclusion, the standard density-density intercomponent interactions are insufficient for a dynamical formation of a long-living dark soliton in the considered setup. 

\begin{figure}[h!]
\includegraphics[width=1.\columnwidth]{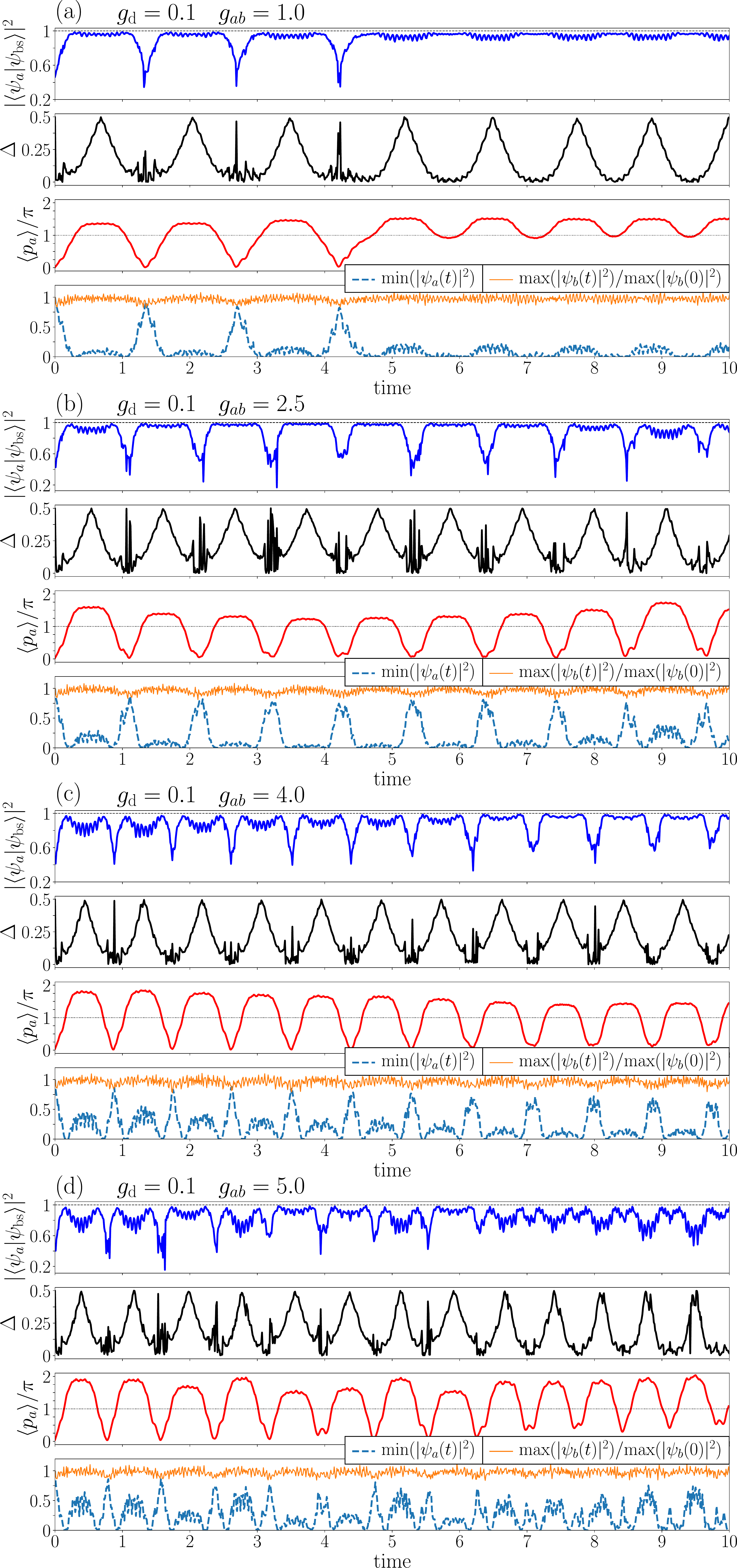}
\vspace{-0.5cm}
\caption{Competition between drag and density-density intercomponent interactions.
From top to bottom the time dependence of: the overlap $|\langle \psi_a|\psi_\mathrm{bs} \rangle|^2$, the relative distance $\Delta$ along the ring between the bright soliton in $b$ component and a phase flip position in $a$ component, the average momentum $\langle p_a\rangle/\pi$, as well as the values $\mathrm{min}|\psi_a(t)|^2$ and $\mathrm{max}|\psi_b(t)|^2/\mathrm{max}|\psi_b(0)|^2$.
The results in (a), (b), (c), and (c) correspond to $g_b=-30$, $g_a=75$, $g_\text{d}=0.1$, and $g_{ab}=1,2.5,4,5$, respectively. 
}
\label{both}
\end{figure}

As we have shown in the main text, the long-living dark soliton can be dynamically induced with the help of drag interactions. That is, while the density-density interactions tend to an abrupt and continuous transfer of momentum between the components which is an impediment for a long-living soliton structures formation, the drag interaction provides a stabilization mechanism supporting a long time existence of a drag-induced dark soliton.     
 Here, we address the question if the effect can still exist when in addition to the drag interactions the density-density  intercomponent interactions are present. In Fig.~\ref{both} we show how the competition between the two types of intercomponent interactions affects the system dynamics for previously discussed parameters $g_a=75$, $g_b=-30$, $g_\mathrm{d}=0.1$, cf., Fig.~3(c) in the main text for the results with exclusively drag interactions. Note that the clearly visible (deep) dark soliton can be still induced and survive for a relatively long time in the evolution even in the presence of not too strong density-density intercomponent interactions. Indeed, while for $g_{ab}=1$ the system dynamics is almost unaffected by density-density intercomponent interactions, one can easily observe shortening the lifetimes of generated dark soliton when increasing $g_{ab}$. Nevertheless, even for $g_{ab}=5$ a quite deep dark soliton survives for time significantly longer than the period of a single revolution of a black soliton around the ring during its drag-free dynamics, i.e., $T=1/\pi$. Not surprisingly, the effect of a long-living dark soliton dynamical formation disappears for stronger density-density intercomponent interactions when they dominate in the system dynamics.

\end{document}